\documentclass[12pt]{article}
\usepackage{amssymb,amsbsy,latexsym}

\newcommand{\ett}{{\mathbf 1}}
\newcommand{\noll}{{\mathbf 0}}
\newcommand{\pL}{\mbox{$\frac{|p|}{\Lambda}$}}
\newcommand{\mL}{\mbox{$\frac{m}{\Lambda}$}}
\newcommand{\id}{I}
\newcommand{\R}{{\mathbb R}}
\newcommand{\C}{{\mathbb C}}

\newcommand{\QED}{\hfill$\square$}

\newcommand{\vs}{\underline{s}}
\newcommand{\vnu}{\underline{\nu}}
\newcommand{\veps}{\underline{\eps}}

\newcommand{\Slog}{S_{\rm log}}
\newcommand{\intS}{\int_{\R^4}}
\newcommand{\intM}{\int_{M^4}}
\newcommand{\intL}{\int_{|p|\leq\Lambda } }
\newcommand{\intML}{\int_{m\leq |p|\leq\Lambda } }
\newcommand{\dvp}{\frac{\dd^{4}p}{(2\pi)^{4} }}
\newcommand{\dvxi}{\frac{\dd^{4}\xi}{(2\pi)^{4} }}
\newcommand{\dvq}{\frac{\dd^{4}q}{(2\pi)^{4} }}
\newcommand{\dvx}{d^{4}x\,}
\newcommand{\dvy}{d^{4}y\,}

\newcommand{\DD}{\hat D}
\newcommand{\Asl}{\hat A}
\newcommand{\Dn}{\DD_0}
\newcommand{\DA}{\DD_A}
\newcommand{\oDD}{\mathsf{D}}
\newcommand{\oAsl}{\mathsf{A}}
\newcommand{\oDn}{{\oDD_0}}
\newcommand{\oDA}{{\oDD_A}}
\newcommand{\oDB}{{\oDD_B}}

\newcommand{\dsl}{\partial \!\!\!\slash}
\newcommand{\psla}{p\!\!\!\slash}
\newcommand{\xxsl}{\tilde{p}\!\!\!\slash}
\newcommand{\asl}{a\!\!\!\slash}
\newcommand{\xisl}{\xi\!\!\!\slash}

\newcommand{\Tr}{{\rm Tr}}          
\newcommand{\TR}{{\rm TR}}          
\newcommand{\TraL}{{\rm Tr}_\Lambda}       
\newcommand{\tra}{{\rm tr}}         
\newcommand{\tras}{{\rm tr}_{\nu}}        
\newcommand{\trac}{{\rm tr}_N}        

\newcommand{\dd}{{\rm d}}
\newcommand{\half}{\mbox{$\frac{1}{2}$} }

\newcommand{\gl}{{\rm gl}}
\newcommand{\GL}{{\rm GL}}
\newcommand{\ee}[1]{\,\mbox{{\rm e}}^{#1}}
\newcommand{\ii}{{\rm i}}

\newcommand{\eps}{\varepsilon}

\newcommand{\cD}{{\cal D}}
\newcommand{\cJ}{{\cal J}}
\newcommand{\cO}{{\cal O}}
\newcommand{\cR}{{\cal R}}
\newcommand{\cP}{{\cal P}}
\newcommand{\cM}{{\cal M}}
\newcommand{\cN}{{\cal N}}
\newcommand{\cH}{{\cal H}}
\newcommand{\cF}{{\cal F}}
\newcommand{\cG}{{\cal G}}
\newcommand{\cI}{{\cal I}}
\newcommand{\cS}{{\cal S}}

\newcommand{\eq}{\begin{equation}}
\newcommand{\eqend}{\end{equation}}
\newcommand{\eqa}{\begin{eqnarray}}
\newcommand{\nonueqa}{\begin{eqnarray*}}
\newcommand{\eqaend}{\end{eqnarray}}
\newcommand{\nonueqaend}{\end{eqnarray*}}
\newcommand{\nonu}{\nonumber \\ \nopagebreak}
\newcommand{\bma}[1]{\begin{array}{#1}}
\newcommand{\ema}{\end{array}}
\newcommand{\bc}{\begin{center}}
\newcommand{\ec}{\end{center}}

\newcommand{\Ref}[1]{(\ref{#1})}

\newcounter{saveeqn}
\newcounter{App} 
\newcommand{\app}{%
\stepcounter{App}%
\setcounter{saveeqn}{\value{equation}}%
\setcounter{equation}{0}%
\renewcommand{\theequation}{\Alph{App}\arabic{equation}} }

\begin{document}
\begin{flushright}
April 8, 2001
\end{flushright}
\vspace{.4cm}

\begin{center}
{\Large \bf Generalized Yang-Mills actions from Dirac operator
determinants}\\

\vspace{1 cm}
{\large Edwin Langmann}\\ 
\vspace{0.3 cm} {\em Theoretical Physics, Royal Institute of
Technology, SE-10044 Stockholm, Sweden}
\vspace{0.3 cm}
\end{center}

\begin{abstract}
We consider the quantum effective action of Dirac fermions on four
dimensional flat Euclidean space coupled to external vector- and axial
Yang-Mills fields, i.e., the logarithm of the (regularized)
determinant of a Dirac operator on flat $\R^4$ twisted by generalized
Yang-Mills fields.  According to physics folklore, the logarithmic
divergent part of this effective action in the pure vector case is
proportional to the Yang-Mills action.  We present an explicit
computation proving this fact, generalized to the chiral case.  We use
an efficient computation method for quantum effective actions which is
based on calculation rules for pseudo-differential operators and which
yields an expansion of the logarithm of Dirac operators in local and
quasi-gauge invariant polynomials of decreasing scaling dimension.

\bigskip 


\noindent MSC-class: 81T13; 58J42; 35S99

\end{abstract}

\section{Introduction}
Determinants of differential operators arise as (exponentials of)
effective actions in quantum field theory.  The precise definition and
investigation of such objects is an interesting and challenging
mathematical problem which has lead to an active and fruitful
interplay between mathematics and physics.

In this paper we compute the logarithmic divergent part, $\Slog(A)$,
of the logarithm of the regularized determinant for Dirac operators
$\oDA$ describing Dirac fermions coupled to a generalized Yang-Mills
field $A$ on four dimensional spacetime. For simplicity we assume
spacetime to be flat $\R^4$ with Euclidean signature and the natural
spin structure.  The Yang-Mills fields we consider contain, besides
the vector part $V$, also a chiral (axial) part $C$ (for precise
definitions see Eq.\ \Ref{16a} {\em ff.} below); we write
$A=(V,C)$. Our definition of $\Slog$ is motivated by physical
considerations and will be explained further below. To indicate the
mathematical significance of our calculation, we note that $\Slog(A)$
is (essentially) the noncommutative residue \cite{W} of the logarithm
of $\oDA$ (see Eq.\ \Ref{Wres} for the precise statement).  A main
motivation for this work is to present a computation method for
effective fermion actions which at the same time is mathematically
rigorous, close to standard Feynman diagram computations in quantum
field theory (see, e.g., \cite{IZ,We}), and simple to use.  We believe that
this method is a useful alternative to other methods like the
$\zeta$-function regularizations or the heat kernel expansions (see,
e.g., \cite{Gil,BGV}).  We therefore made some effort to present this
method in a self-contained way, in the hope that this is useful also
for readers who are mainly interested in learning how to compute
effective actions.

We now discuss our computation method (parts of this method were used
previously by us in \cite{L1,LM1}). We regard the Dirac operator $\oDA$
as a PSDO (pseudo-differential operator) on a Hilbert space of
square-integrable functions on $\R^4$.  Our starting point is the
following definition for the regularized effective fermion action,
\eq
\label{Seff}
S_{\Lambda}(A) :\; = \TraL \left( \log \left(\frac{\oDA+\ii m
}{\Lambda_0}\right) - \log \left(\frac{\oDn+\ii m}{\Lambda_0}\right)
\right) 
\eqend 
where $m$ is a real parameter which has the physical interpretation of
a fermion mass, and $\Lambda$ is a positive regularization parameter
which we call {\em UV} (ultra-violet) {\em cutoff}.  The role of the
non-zero and complex parameter $\Lambda_0$ is two-fold.  Firstly, it
makes the argument of the logarithm dimensionless, and secondly,
setting $\Lambda_0=|\Lambda_0|/(1+\ii 0^+)$ avoids possible
ambiguities due to the branch cuts of the logarithm which otherwise
can arise.\footnote{Of course, all results must be independent of
$|\Lambda_0|$, and this is a useful check.}  This definition above has
three ingredients. Firstly, a definition of the $\log$ of an operator
$a$ as an integral of the resolvent of $a$. Secondly, some basic facts
about PSDO which imply a simple and powerful formula for the symbol of
the resolvent of the Dirac operator $\oDA$.  And thirdly, a definition
of a regularized Hilbert space trace $\TraL$ (where removing the
regularization corresponds to the limit $\Lambda\to
\infty$). Combining these ingredients we obtain an expansion of
$S_{\Lambda}(A)$ in local and quasi-gauge invariant polynomials of
decreasing scaling dimension. We find
\eq
\label{expand} 
S_{\Lambda}(A) =
\Lambda^2 S^{(2)}(A) + \log\left(\frac{\Lambda}{|m|}\right) \Slog(A) +
S^{(0)}(A) + \cO(\Lambda^{-1}) , \eqend
and this provides our definition of $\Slog(A)$. Our results for
$\Slog(A)$ and $S^{(2)}(A)$ will be presented in the next Section.  
We shall also demonstrate on our way that $\Slog(A)$ is
proportional to the noncommutative residue \cite{W} of the logarithm
of the Dirac operator $\oDA$,
\eq
\label{Wres}
\Slog(A) = 4 \, {\rm Res}\left( \log \left(\frac{\oDA+\ii m
}{\Lambda_0}\right) - \log \left(\frac{\oDn+\ii m}{\Lambda_0}\right)
\right) \: .  \eqend The logarithm of the regularized trace of the
determinant of the Dirac operator can then be defined as
\eq 
\label{S0}
S^{(0)}(A) =
\TR \left( \log \left(\frac{\oDA+\ii m }{\Lambda_0}\right) - \log
\left(\frac{\oDn+\ii m}{\Lambda_0}\right) \right)
\eqend 
where $\TR$ is the renormalized trace which we will define, and we
will provide all mathematical tools necessary for computing
$S^{(0)}(A)$ explicitly.

We note that our computation method is closely related to methods
which have been used in the physics literature for a long time (see,
e.g., \cite{dW,IZ,We}).  The regularization we use is simple and
close to how regularizations are often done in Feynman diagram
computations, i.e., by introducing a sharp UV cutoff (see Eq.\ 
\Ref{TrLam}).  We believe, however, that we can offer some
improvements in detail which make computations easier, more
transparent in structure, but nevertheless such that each step
can be easily justified with mathematical rigor.

We now discuss some motivation for our computation from a quantum
field theory point of view.  As was known already to Schwinger for the
Abelian case, the effective action of fermions coupled to a Yang-Mills
field $A=V$ (i.e., $C=0$) contains a logarithmic divergence,
$\log(\Lambda/m)\, \Slog(A)$, and $\Slog(A)$ (for $C=0$) is proportional
to the usual Yang-Mills action 
$$
S_{\rm YM}(A) = \frac{1}{ 2 g^2}\, \intS   \dd^4
x\; {\rm tr}\, \cF_{\mu\nu}\cF^{\mu\nu}
$$
(see, e.g., \cite{IZ}, Eq.\ (12.123) where $1/\epsilon$ corresponds to
$\log(\Lambda/m)$).  This is important since it implies that a change
in the cutoff in the gauge theory, $\Lambda\to \Lambda'$, leads to a
finite change of the effective fermion action which can be absorbed by
changing the Yang-Mills coupled constant, $g^{-2}\to (g')^{-2} =
g^{-2} + const.  \log(\Lambda' / \Lambda)$. The logarithmic dependence
of the Yang-Mills coupling constant on the UV cutoff is remarkable and
distinguishes four spacetime dimensions from all others.

Our computation is closely related to more recent ideas which have
lead to a deeper geometric understanding of the standard model of
elementary particle physics (including Higgs sector).  This approach
is based on Connes' NCG (noncommutative geometry; textbooks on this
subject are, e.g., \cite{NCG,GVF}).  One important ingredient of this
approach is to define a generalized Dirac operator $\oDA$, and this
Dirac operator not only specifies the fermion part of the action of
the model but also the Yang-Mills part $S_{YM}(A)$: there is a
definition of $S_{YM}(A)$ in terms of $\oDA$ (see \cite{CC} and
references therein). Our discussion above suggests a simple physical
interpretation of this {\em spectral action principle} \cite{CC}: the
logarithmic divergence of the fermion effective action is potentially
`dangerous' since it can make the model ambiguous: there is no
preferred choice for the cut-off, and changing it generates a term
proportional to $S_{\rm log}(A)$. However, the fact that $\Slog(A)$ is
proportional to the Yang-Mills action resolves this problem for the
standard (purely vector) Yang-Mills theory on $\R^4$, as discussed
above. It therefore is natural to {\em require that the Yang-Mills
action is proportional to the logarithmic divergent part of the
fermion effective action in any gauge theory models}. In
particular this suggests the following definition of the generalized
(vector and chiral) Yang-Mills action in terms of the generalized
Dirac operators $\oDA$,
\eq
\label{NCAC}
S_{YM}(A) := const.\,  \frac{1}{ 2 g^2}\, \Slog(A)   \eqend 
(for one fermion flavor $const. = 24\pi^2$).  Eq.\ \Ref{Wres} shows
that for flat Euclidean space $\R^4$, this definition is equivalent to
the one given in \cite{CC}. We conjecture that this is true for other
four dimensional spin manifolds as well.

The plan of this paper is as follows. We summarize our notation and
results in Section 2. Section 3 contains a summary of the mathematical
prerequisites, i.e., the three ingredients of our method mentioned
above. The computations of $\Slog(A)$ is presented in Section 4 with
some computation details deferred to Appendix B.  We conclude with
some remarks in Section 5. Appendix A contains some discussion on
regularized traces and the noncommutative residue.

\bigskip
\noindent {\bf Notation:} We write $\gl_N$ for the complex $N\times N$
matrices and $\GL_N$ for the invertible matrices in $\gl_N$. We
sometimes write $\id_V$ or $\id$ for the identity operator on a vector
space $V$ but often abuse notation and do not distinguish between $c
\id$ and $c$ for complex numbers. For $V$, $W$ vector spaces and $a$
an operator on $V$, we often use the same symbols $a$ to also denote
the corresponding operator $a \otimes \id_W$ and $\id_W\otimes a$ on
$V\otimes W$ and $W\otimes V$, respectively.  The real part of a
complex number $c$ is denoted as $\Re c$.

\section{Definitions and Results}
\label{sec2}
For simplicity we assume spacetime $M^4=\R^{4}$ with Euclidean
signature (the extension of our calculation to other four--dimensional
spin manifolds should be possible using symbol calculus of
pseudo--differential operators \cite{Hormander}).

We consider the Hilbert \eq \cH=L^2(\R^4)\otimes\C^4_{spin}
\otimes\C^N_{color} \eqend which has the physical interpretation as
space of the 1--particle states of the fermions. We also introduce the
space $\cD$ of functions in $\cH$ which are smooth (i.e., $C^\infty$)
and $L^1$; $\cD$ is a convenient dense domain in $\cH$.

The Dirac operators of interest to us are of the form
\eq
\label{16a}
\DA = \gamma^\nu\left( -\ii \partial_\nu + V_\nu(x) + \ii \gamma_5
C_\nu(x) \right) 
\eqend 
where $A=(V,C)$ (repeated indices $\nu,\mu\ldots =1,2,3,4$ are summed
over; $x=(x^1,x^2,x^3,x^4)\in \R^4$), with
$\partial_\nu=\frac{\partial}{\partial x^\nu}$ and $\gamma^\nu$ the Dirac
spin matrices acting on $\C^4_{spin}$ and obeying 
\eq
\label{16b}
\gamma^\mu\gamma^\nu +\gamma^\nu\gamma^\mu = 2\eta^{\mu\nu}
\eqend
for $\mu,\nu=1,2,3,4$, where $ \eta^{\mu\nu}=\eta_{\mu\nu} = diag(1,
1, 1, 1) $ is the metric tensor, and 
\eq 
\gamma_5:\; =\gamma^1\gamma^2\gamma^3\gamma^4 
\eqend
as usual (for the convenience of the reader, explicit formulas for
these matrices are given in Appendix~A.1).

For simplicity we assume that the
functions $V_\nu$ and $C_\nu$ $\R^4\to \gl_N$ are {\em regular}, i.e.,
they are $C^\infty$ and vanish like $\cO(|x|^{-4-\eps})$, for some
$\eps>0$, as $|x|\to \infty$ (the latter condition is to ensure that
integrals of regular functions over $\R^4$ absolutely converge).

In particular, the free Dirac operator is defined by the differential
operator 
\eq
\label{160}
\Dn= -\ii \gamma^\nu\partial_\nu \: . 
\eqend

We define the {\em gauge group} $\cG$ as follows. Let $\GL_N$ be the
group of all invertible matrices in $\gl_N$. Then $\cG$ is the group
of all $\GL_N$-valued functions $U$ on $\R^4$ such that $U(x)-1$ is a
regular function.  Note that one can write
$$
\DA= \half(1-\gamma_5) \gamma^\nu \left(-\ii\partial_\mu + V_\mu + \ii
C_\mu \right) + \half (1+\gamma_5) \gamma^\nu \left(-\ii\partial_\mu +
V_\mu - \ii C_\mu \right)
$$
where $ V_\mu \pm \ii C_\mu $ are the {\em chiral} components of the
gauge field. This representation shows that it is natural to consider
two kinds of gauge transformations, 
\eq
\label{gt}
V_\mu \pm \ii C_\mu   \to (U_\pm)^{-1} \left(V_\mu \pm \ii C_\mu 
 \right) U_\pm  -\ii 
(U_\pm)^{-1} \partial_\mu U_\pm, \quad U_\pm\in \cG \: . 
\eqend
For $U_+=U_-=U$ we denote these as {\em vector gauge transformation},
otherwise as {\em chiral gauge transformation}.

Note that $\DA$ in Eq.\ \Ref{16a} is well-defined on the domain
$\cD\subset\cH$, and we find it useful to distinguish this formally
self-adjoint differential operator in notation from the corresponding
self-adjoint extension on $\cH$ which we denote as
$\oDA$, i.e., $(\oDA f)(x)=\DA f(x)$ for all $f(x)\in\cD$.  
We also write 
\eq \oDA =\oDn+\oAsl \eqend where $\oDn$ is the free Dirac operator
(i.e., self-adjoint extension of $\Dn$) and $\oAsl$ the operator
defined by multiplication with the generalized Yang-Mills field
\eq \Asl(x)=\sum_{\nu=1}^4 \gamma^\nu \left( V_{\nu} (x) + \ii
\gamma_5 C_\nu(x) \right) \: .  
\eqend

We will compute the fermion effective action $S_{\Lambda}(A)$ defined
in Eq.\ \Ref{Seff}, and we will show that it can be expanded as in
Eq.\ \Ref{expand}.  As discussed, $\TraL$ is a Hilbert space trace
with an ultraviolet (UV) cutoff $\Lambda>0$, and $\Lambda_0$ is an
arbitrary, in general complex, parameter makes the argument of the
logarithm dimensionless. Moreover, the real (positive or negative)
parameter $m$ corresponds to a fermion mass and serves as an infrared
(IR) regulator in our computation.  Our main result is an explicit
formula for $\Slog(A)$.

\bigskip

\noindent {\bf Proposition:} {\it The logarithmic divergent piece
$\Slog(A)$ of the logarithm of the (regularized) determinant of the
Dirac operator $\oDA$ equals
\eq
\label{main}
\Slog(A) = \frac{1}{24\pi^2}  \intM\dvx\trac\left( 
\half \cF^+_{\mu\nu} (\cF^+)^{\mu\nu} + \half \cF^-_{\mu\nu} (\cF^-)^{\mu\nu} 
- 6 m^2 C^\mu C_\mu \right)
\eqend
where $\trac$ is the usual 
matrix trace in $\gl_N$ and\footnote{$[a,b]:=ab-ba$}  
\eq 
\label{cFpm}
\cF^\pm_{\mu\nu}\, :=\, \partial_\mu A^{\pm}_\nu - \partial_\nu
A^{\pm}_\mu + \ii[A^{\pm}_\mu,A^{\pm}_\nu], \quad A_\mu^\pm \, :=
\,V_\mu\pm C_\mu
\eqend
is the curvature associated with the chiral component $A^{\pm}$ of the
Yang-Mills field.}

\bigskip

\noindent {\em (Proof in Section~\ref{sec4} with some details deferred
to Appendix~B.) }

\bigskip

For $C=0$ (no chiral field) we obtain 
$$
\Slog(A) =  \frac{1}{24\pi^2}\intM\dvx\trac \cF_{\mu\nu} \cF^{\mu\nu} 
$$
with 
\eqa
\cF_{\mu\nu}= \partial_\mu V_\nu - \partial_\nu V_\mu + \ii[V_\mu,V_\nu], 
\eqaend
which is the standard Yang-Mills action.  Note that $\cF_{\mu\nu}=\ii
[D_\mu,D_\nu]$ with
\eq
D_\nu := -\ii \partial_\nu  + V_\nu (x) 
\eqend
the covariant derivative, and similarly, 
\eq
\cF^\pm_{\mu\nu} = \ii [D_\mu \pm \ii C_\mu, D_\nu \pm \ii C_\nu] . 
\eqend

It is important to note that for $m=0$, $\Slog(A)$ in Eq.\  \Ref{main} 
this is manifestly invariant under all gauge 
transformations Eq.\  \Ref{gt}. For $m\neq 0$, 
there is also a mass term
$\propto C^\mu C_\mu$ for the chiral gauge field which is 
only invariant under vector gauge transformations, 
i.e., only the transformations Eq.\  \Ref{gt} with $U_+=U_-=U$.  
The parameter in front of this term is fixed by the fermion mass. 
There is no similar term for 
the vector gauge field (note that such a term
would spoil vector gauge invariance).

It is interesting to note that the result of our computation in
Section \ref{sec4} suggests that for manifolds $M^4$ with boundary
$\partial M^4$, $\Slog(A)$ has an additional contribution 
\eq
\label{main1}
\Delta\Slog(A) = \frac{1}{24\pi^2 }\intM\dvx \partial^\mu \trac 
J_\mu,
\eqend
with 
\eq
\label{J}
J_\mu : =  2 C_\mu\ii [D_\nu , C^\nu] -2 C^\nu \ii [D_\nu , C_\mu] 
+ 2 \ii [D_\mu,C^\nu C_\nu ] \; . 
\eqend
This is a boundary term (by Stokes's theorem). Note that this term is also 
invariant under vector gauge transformations, 
and it vanishes if the axial Yang-Mills field $C_\mu$ is zero.

It is also worth noting that, as a by-product, we also obtain the
explicit expression for the quadratic divergent part of the effective
action, \eq
\label{S2}
S^{(2)}(A) = \frac{1}{16 \pi^2}
\intM \dvx \trac(-V^\mu V_\mu + C^\mu C_\mu ) \: . 
\eqend
In contrast to $\Slog(A)$ this term is not gauge invariant (as already
mentioned, the term $\propto V^\mu V_\mu$ spoils vector gauge
invariance)!  This highlights the fact that the regularization
procedure we use it not manifestly gauge invariant but only
quasi-gauge invariant.  It shows that the vector gauge invariance of
our result for $\Slog$ somewhat remarkable. It is also interesting to
note that for $V_\mu= \pm C_\mu$, $S^{(2)}(A)=0$.

\section{Calculation tools}\label{sec3}
In this Section we collect the mathematical prerequisites for our 
computation.  
We will explain the three ingredients for our method: 
Firstly, a definition of the 
logarithm of operators $a$ in terms of an integral of the resolvent 
of $a$. 
Secondly, a few basic definitions for PSDO which imply a simple 
and elegant formula for the symbol of the resolvent of Dirac 
operators $\oDA$. And finally, a definition of a regularized Hilbert space trace 
$\TraL$ (corresponding to introducing an UV cutoff $\Lambda$).  
In the next Section we will put these ingredients together and 
obtain an expansion of the effective action as described in the 
Introduction. 

\bigskip
\noindent
{\bf 1. The logarithm of operators.} 
Let $a$ be a bounded operator on a Hilbert space $\cH$ with norm less 
then one. Then ($\id=\id_{\cH}$ is the identity operator) 
\eq
\label{log}
\log (\id +a) = \int_0^1 \frac{\dd s}{s} \left( \id - 
(\id +s a)^{-1} \right),  
\eqend
as can be seen by a Taylor expansion,
$$
\log (\id +a) = \sum_{n=1}^\infty \frac{(-1)^{n-1}}{n}a^n = 
-\sum_{n=1}^\infty \int_0^1 \frac{\dd s}{s}  (-s a)^{n}, 
$$
interchanging summation and integration, and using the geometric series. 

We take this as a motivation to {\em define}
\eq
\label{logD}
\log\left( \frac{\oDA +\ii m }{\Lambda_0} \right) :\; = \int_0^1 \frac{\dd
s}{s} \left( \id - \left(\id+s \left[\frac{\oDA+ \ii m }{\Lambda_0}-\id
\right] \right)^{-1} \right) \eqend where $\Lambda_0$ is a some
complex number. This representation of the logarithm as integral of a
resolvent will be convenient for us since there is a simple formula
for the resolvent of (generalized) Dirac operators, as discussed
below.

\bigskip
\noindent
{\bf 2.A. Pseudo--differential operators. Generalities.}  We summarize
some basic facts about pseudo--differential operators (PSDO) on $\R^4$
(a discussion for general manifolds can be found, e.g., in 
\cite{Hormander}). We consider PSDO $a$ on $\cH$ which can be
represented by their {\em symbol} $\sigma[a](p,x)$, i.e., a $\gl_4
\otimes \gl_N$--valued functions on phase space $\R^{4}\times\R^{4}$
defined such that \cite{Hormander} \eq
\label{s}
(af)(x) = \intS\dvp \intS \dvy\, e^{ip\cdot(x-y) }
\sigma[a](p,x) f(y) 
\eqend
for all $f(y)\in \cD$ (matrix multiplication is understood; 
$p \cdot x= x^\nu p_\nu$).  In particular, $\oDn$ and $\oAsl$ are PSDO with 
symbols 
\eq
\sigma[\oDn](p,x)=\psla:=\gamma^\nu p_\nu, \quad 
\sigma[\oAsl](p,x)=\Asl(x)\: .
\eqend

Note that Eq.\  \Ref{s} implies the following equation  which 
encodes the product of operators in terms of their symbols, 
\eq
\label{product}
\sigma[ab](p,x)= \intS\dvq\intS\dvy\, \ee{\ii(x-y)\cdot(p-q)}
\sigma[a](q,x)\sigma[b](p,y). 
\eqend

We will encounter PSDO $a$ 
which allow an asymptotic expansion
\eq
\label{as0}
\sigma[a]\sim\sum_{j=0}^{\infty}\sigma_{K-j}[a]
\eqend
where $\sigma_{K-j}[a](p,x)$ is
homogeneous of degree $K-j$ in
$p$,\footnote{i.e., $\sigma_{K-j}[a](s p,x)=s^{K-j}\sigma_{K-j}[a](p, x)$ for
all $s>0$ and $|p|>0$} and goes to zero like $|p|^{K-j}$ for $|p|\to\infty$ 
($|p| :\, =\sqrt{p\cdot p}$).  We write
\eq
\sigma[a](p,x)=\sum_{j=0}^{n}\sigma_{K-j}[a](p,x) + \cO(|p|^{K-n-1})
\eqend
for all integers $n$. Eq.\  \Ref{product} implies,
\eq
\label{as}
\sigma[ab](p,x) \sim \sum_{n=0}^{\infty} \frac{(-\ii)^n}{n!}
\frac{\partial^n\sigma[a](p,x)}{\partial p_{i_1}\cdots\partial p_{i_n}}
\frac{\partial^n\sigma[b](p,x)}{\partial x_{i_1}\cdots\partial x_{i_n}}.
\eqend
This equation allows to determine the asymptotic expansions of $\sigma[ab]$ 
and $\sigma[a^{-1}]$ from the ones of $\sigma[a]$ and $\sigma[b]$.  

\bigskip
\noindent
{\bf 2.B. The symbol of the resolvent.}  Eq.\  \Ref{logD} expresses
$\log(\oDA+\ii m)$ as an integral of resolvents of the Dirac operator
$\oDA$, i.e., of operators $\left(c_1\id + c_2\oDA \right)^{-1}$ with
$c_{1,2}$ complex numbers. We will therefore need the symbol of such a 
resolvent. To determine this we note that \eq
\label{hg}
\sigma[c_1 \id + c_2 \oDA ](p,x) =c_1 + c_2[ \psla + \Asl(x) ] \: .
\eqend 
We then could use Eq.\  \Ref{as} to find the expansion
for $\sigma[\left(c_1\id + c_2\oDA \right)^{-1} ](p,x)$.
We now present a useful result summarizing this expansion in a simple
formula.
 
\bigskip
\noindent {\bf Lemma:} {\it The following holds for all
$c_1,c_2\in\C$, \eq
\label{basic}
\sigma[\left(c_1\id + c_2\oDA \right)^{-1} a ](p,x) = 
\left( c_1 +c_2[ \psla +\DA ]   
\right)^{-1}\sigma[a](x,p) .  
\eqend
} 

\bigskip
\noindent {\em Remark:} 
The proper interpretation of this equation is as follows,
$$
\sigma[\left(c_1\id + c_2\oDA \right)^{-1} a ](p,x) \sim 
\sum_{n=0}^\infty (-1)^n \left( c_1 +c_2\psla \right)^{-1}\left[  
\DA\left( c_1 +c_2\psla \right)^{-1}\right]^n \sigma[a](x,p) 
$$
where the differential operators $\partial_\nu$ in 
$\DA=-\ii\gamma^\mu \partial_\nu+\Asl(x)$ act to the right 
on the functions $\Asl(x)$ according to the Leibniz rule. 
We note that we will need this equation only for $a=\id$.

\bigskip
\noindent {\em Proof of the Lemma:} One can check Eq.\  \Ref{basic}
by using Eqs.\  \Ref{as0} and \Ref{as}, taking $c_1 \id +c_2 \oDA$
for $a$ and $[c_1 \id +c_2 \oDA]^{-1}a$ for $b$, and inserting
Eq.\  \Ref{hg}. A simpler argument avoiding tedious expansions is
as follows: Note that by definition, $(\oDA f)(x)=[\Dn+\Asl(x)] f(x)$
for all $f\in\cD$, thus \nonueqa \left( \left( c_1 \id +c_2 \oDA
\right) af\right)(x) = \left(c_1+ c_2 [ \Dn +\Asl(x) ] \right)(af)(x)
= \\ \intS\dvp \intS \dvy \, e^{ip\cdot (x-y) } \left(c_1+ c_2 [ \psla
+\Dn +\Asl(x) ] \right) \sigma[a](p,x) f(y) \nonueqaend where we used
Eq.\  \Ref{s} and the Leibniz rule.  Replacing $a$ in this equation  
by $\left(c_1\id + c_2\oDA \right)^{-1} a$, we see that this is
equivalent to Eq.\  \Ref{basic}. (Note that this argument implies
the interpretation of Eq.\  \Ref{basic} as given above!)  \QED

\bigskip
\noindent {\em Remark:}
We believe that our  expansion in powers of the differential operator 
$\DA$  is very natural for 
at least two reasons. Firstly, since under a vector gauge transformation, 
$\DA \to U^{-1}\DA U$, such an expansion 
is close to being manifestly gauge invariant (we will discuss this 
point in more detail below). Secondly, it is natural from the point of 
view of power counting: in contrast to an expansion in $\Asl$,
the $n$-th order term in our expansion includes precisely those 
local polynomials $\cP_n$ in $V_\mu$ and $C_\nu$ (and derivatives thereof) 
which all have the same scaling 
behavior $\cP_n\to \lambda^{4-n}\cP_n$ under $x\to \lambda x$. 

\bigskip
\noindent {\em Remark:} Loosely speaking, PSDO are useful since they
allow to interpolate between Fourier- and position space: generically
in quantum theory one deals with operators $H$ on some Hilbert
space of $L^2$-functions on $\R^n$ which are a sum of a free part
$H_0$ diagonal in Fourier space,\footnote{$\hat f(p)= \int_{\R^n}
\dd^n x\, e^{ip\cdot x}f(x)$ denotes the Fourier transform.}
$\widehat{(H_0f)}(p)=E_0(p)\hat f(p)$, and a potential term $V$
diagonal in position space, $(Vf)(x)=V(x)f(x)$. The symbol of
$\sigma[H](p,x)$ is then simply the sum of $E_0(p)$ and $V(x)$, which
is an attractive feature. The price one has to pay is that the symbol
of (`nice') functions $F$ of $H$ are somewhat complicated: in a first
approximation, $\sigma[F(H)](p,x)\sim F(E_0(p)+V(x)) + \ldots$, but
there are correction terms $\ldots$ depending on derivatives. The
Lemma above is a special case of the following formula,
$$
\sigma[F(H)](p,x)\sim F(E_0(p-\ii\partial)+V(x)) 1
$$
nicely summarizing the systematic derivative expansion of functions of
$H$.

\bigskip
\noindent
{\bf 3. Regularized traces and the noncommutative residue.}  
We now define the regularized trace which we will use.  We first note
that due to our technical assumptions on the gauge fields all
operators $a$ which we will encounter are PSDO which have symbols
$\sigma[a](p,x)$ which go at least like $\cO(|x|^{-4-\eps})$, some
$\eps>0$, for fixed $p$ and $|x|\to\infty$, and are finite for finite
$p$. Thus 
\eq
\label{TrLam}
\TraL(a) :\; = \intL \dvp \intS \dvx\, \tra\, \sigma[a](x,p)
\eqend 
where $\tra$ is the full matrix trace,\footnote{including the trace
$\tras$ in $\gl_4$ and the trace $\trac$ in $\gl_N$.}  is well-defined
for $\Lambda<\infty$, and this defines a regularized Hilbert space
trace: If $a$ is a trace--class operator then $\TraL(a)$ has a
well--defined limit $\Lambda\to\infty$ which is equal to the Hilbert
space trace of $a$ \cite{Hormander}.  More generally one can consider
PSDO $a$ for which $\TraL(a)$ can be expanded as
\eqa 
\label{TrReg}
\TraL(a) = c^{(K)}(a)\Lambda^K + c^{(K-1)}(a)
\Lambda^{K-1} +\nonu + \ldots + c^{(1)}(a)\Lambda + c_{\log}(a)
\log\left(\frac{\Lambda}{|m|}\right) + c^{(0)}(a) + {\cal O}(\Lambda^{-1})
\eqaend 
with $K$ some non--negative integer.

We recall that the noncommutative residue \cite{W} of a PSDO $a$ with
an asymptotic expansion as in Eq.\  \Ref{as0} can be defined as
(see, e.g., Eq.\ (2.7) in Ref.\ \cite{VGB})
\eq 
\label{Wres_def}
{\rm Res}(a)\, : = \frac{1}{4}\intS\dvp\, \delta(|p|-1)\intS\dvx
\tra \sigma_{-4}[a](x,p), 
\eqend
and for PSDO $a$ as above,
\eq
\label{aa}
{\rm Res}(a) = \frac{1}{4} c_{\log}(a), \eqend
i.e., the residue is equal, up to a constant, to the logarithmic
divergent part of the regularized trace of $a$. (An elementary 
proof of this latter fact is outlined in Appendix~A.)

\bigskip
\noindent
{\em Remark:} In our definition Eq.\  \Ref{trr} of $\TraL$ we use a
sharp cutoff, i.e., 
\eq
\label{TrLam1}
\TraL(a) = \intS \dvp f(\pL) 
\intL \dvx\, \tra \sigma[a](x,p) 
\eqend 
where $f(t)$ equals the Heaviside step function
$\theta(1-t)$. In principle one could define a regularized
trace using Eq.\ \Ref{TrLam1} and choosing any non-negative, piece-wise
smooth, function $f(t)$ which vanishes exponentially fast for
$|t|\to\infty$ and is such that $f(0)=1$. For example, the choice
$f(t)=\exp(-t^2)$ would correspond to the standard heat kernel
regularization.

We will show in Appendix A that $c_{\log}(a)$ is in fact independent of $f$.

Using any such regularization one can define the {\em renormalized
trace} as the finite part of the regularized trace,
\eq \TR(a) \, := \,c_0(x) ,   
\eqend
but this is not quite independent of the regularization: as also
discussed in Appendix A, changing the regularization function $f\to
\tilde f$ amounts to changing
\eq 
\label{change}
\TR(a) \to \TR(a) + \log(s) \, c_{\log}(a) 
\eqend 
with some constant $s>0$ depending on $f$ and $\tilde f$: the
logarithmic divergent piece accounts for the regularization dependence
of the renormalized trace, and this is the reason for our interest in
it, as discussed in the Introduction.

\bigskip
\noindent
{\em Remark:} We note Eq.\  \Ref{TrLam1} is equivalent to 
\eq
\label{trr}
\TraL(a)=\Tr(P_\Lambda a), \quad P_\Lambda\, := f( |\oDn|/\Lambda )
\eqend (using the spectral theorem for self-adjoint operators). This
naturally extends the definition of $\TraL$ from PSDO to a large class
of operators on $\cH$.  More generally, one could change the
regularization by changing $\oDn \to \oDB$ in the definition of $
P_\Lambda$, for some fixed Yang-Mills field $B$. One can show that
this would change $\TR(a)$ by a term proportional to ${\rm
Res}([\log(\oDB)-\log(\oDn)]a )$ (see, e.g., Eq.\ (1.6) in
\cite{CDMP}). It would be interesting to explore this possibility in
more detail.

\section{Computation of effective fermion action}\label{sec4}
In this Section we present the explicit computation of the effective
fermion action and thus prove the proposition in Section
\ref{sec2}. Our computation amounts to a quasi-gauge invariant
gradient expansion, which is essentially an expansion
in powers of the UV cutoff $\Lambda$. This allows us to extract, in a
simple manner, the quadratic and logarithmic divergent pieces which is
what we are interested in.

\bigskip
\noindent
{\bf 1. Quasi-gauge covariant expansion.} 
We write 
\eq
\label{one}
S_{\Lambda}(A) = \intL\dvp\intS\dvx\tra \, \cS(x,p) \, 1
\eqend
where $\cS(x,p)$ is obtained by computing the symbol of the operator 
$\log(\oDA+\ii m)/\Lambda_0-\log(\oDn+\ii m)/\Lambda_0$ as explained in 
the last Section, i.e.,  
\eqa
\cS(x,p) = 
\int_0^1 \frac{\dd s}{s} \left( [1 -s + s \xxsl + s \asl ]^{-1} 
- [1 -s + s \xxsl ]^{-1}  \right) =\nonu
\int_0^\infty \frac{\dd u}{u} \left( [1 + u \xxsl + u \asl ]^{-1} 
- [1+ u \xxsl ]^{-1}  \right); 
\eqaend
we used Eqs.\  \Ref{log} and \Ref{basic}, introduced the 
convenient short-hand notion, 
\eq
\xxsl:\; = \frac{\psla + \ii m }{\Lambda_0},\quad \asl:\;= 
\frac{-\ii\dsl + \Asl}{\Lambda_0} ,
\eqend
and changed integration variables, $s=u/(1+u)$.  The $1$ on the
r.h.s.\ of Eq.\ \Ref{one} is the symbol of the identity operator.  As
explained in more detail below, $\cS$ here is to be regarded as a
differential operators acting on $1$. It is straightforward to expand
the integrand in this equation in powers of $\asl$,
\eq
\cS = \sum_{n=1}^L (-1)^{n-1} \cS_n  + \cR_{L+1}  
\eqend
where 
\eq
\cS_n = (-1)^{n-1} \int_0^\infty \frac{\dd u}{u} 
\frac{1}{1+u\xxsl}\left( u \asl \frac{1}{1+u\xxsl}\right)^n
\eqend
and 
\eq
\cR_{L+1} =  \int_0^\infty \frac{\dd u}{u} 
\frac{1}{1+u\xxsl}\left( u \asl \frac{1}{1+u\xxsl}\right)^L 
u \asl [1+u(\xxsl+\asl)]^{-1}
\eqend
is a remainder term. 

In the following 
we find it convenient to use the short-hand notation and write 
\eq
\DA= \sum_{s=0,5} \gamma^\nu_s 
D^{s}_{\nu}
\eqend
where 
\eq
D^0_{\nu}:=D_{\nu} \quad  D^5_{\nu}:=C_{\nu} 
\eqend
and 
\eq
\gamma_0^\nu: =\gamma^\nu, \quad \gamma_5^\nu:=\ii\gamma^\nu\gamma_5 . 
\eqend
We then define
\eq
\label{cM}
\cM^{\nu_1\ldots\nu_n}_{n;s_1\ldots s_n} :=\, (\Lambda_0)^{-n}
\intL\dvp \tras \int_0^\infty \dd u u^{n-1}   \frac{1}{1+u\xxsl}
\gamma_{s_1}^{\nu_1} \frac{1}{1+u\xxsl}\cdots
\gamma_{s_n}^{\nu_n}  \frac{1}{1+u\xxsl}
\eqend
where $s_j=0,5$ and $\nu_j=1,2,3,4$. This allows us to write
\eq
\label{ex}
S_n:= \intL\dvp\intS\dvx\tra \, \cS_n(x,p)  = \sum_{\vs}
\cM^{\nu_1\ldots\nu_n}_{n;\vs} \intS\dvx\trac D^{s_1}_{\nu_1}\cdots
D^{s_n}_{\nu_n}; \eqend
here and in the following, $\vs$ is short for $(s_1,\ldots,s_n)$.

The following Lemma simplifies the computation significantly: it
implies that the $S_n$ for {\em odd} integers $n$ all vanish, and that
an series expansion in the mass $m$ only has non-zero {\em even}
powers.

\bigskip
\noindent {\bf Lemma:} {\it The coefficients $\cM^{\vnu}_{n;\vs}$ in
Eq.\ \Ref{cM} are non-zero only for even integers $n$, and they are
invariant under $m\to -m$, i.e., they are independent of the sign of
the mass.}
\bigskip

\noindent {\em (Proof in Appendix B.)} 

\bigskip
\noindent {\em Remark:} 
We now can explain why we denote our expansion quasi-gauge invariant. 
This is because the operators $D_\nu^s$ transform gauge covariantly 
under a vector gauge transformation $U$, $D_\nu^s\to U^{-1} D_\nu^s 
U$. This implies that the differential operators defined in Eq.\  \Ref{ex} all 
are gauge invariant. However, the action is a polynomial which is 
obtained by applying these differentiation operators to $1$ (cf.\ Eqs.\  
\Ref{one}) using Leibniz rule and $\partial_\nu 1=0$, e.g.,  
\eqa
\label{how}
D_{\nu} 1 &=& V_{\nu}(x)\nonu
D_{\nu_1}C_{\nu_2}1 &=& -\ii[\partial_{\nu_1} 
C_{\nu_2}(x)]  + 
V_{\nu_1}(x)C_{\nu_2}(x)\nonu
D_{\nu_1}D_{\nu_2}1 &=& -\ii[\partial_{\nu_1} 
V_{\nu_2}(x)]  + 
V_{\nu_1}(x)V_{\nu_2}(x)
\eqaend
etc. The result will be gauge invariant only if the differential
operator in Eq.\ \Ref{ex} is already a polynomial. This happens, e.g.,
if the differential operators $D_\nu$ only appear in combinations
$[D_\nu,D_\mu]$ and $[D_\nu,C_\mu]$. This is not obvious.  However, we
will see below that this happens for the terms leading to logarithmic
divergent part.

\bigskip
\noindent
{\bf 2. Expansion in powers of the UV cutoff.}  We now show that our
expansion above is essentially an expansion in powers of the UV cutoff
$\Lambda$.  Our computation can be simplified by the following
argument (this argument is refined and justified in detail in
Appendix~B).  As mentioned, $m$ serves as a particular IR cutoff for
momentum integrals.  We expect that our result is independent of the
precise form the IR regularization. Thus we use instead the following,
simpler one: we set $m=0$ in $\xxsl$ but restrict integrations over
$p$ to $m\leq|p|\leq\Lambda$.  {\em We stress that we use this
simplification in the main text only to ease our presentation, and
that it is appropriate only for computing the diverging contributions
to the regularized determinant: the computation of the finite part
should be done with the method explained in Appendix~B}.  Below we
shall see that this simplified procedure gives a IR regularization
provided we also set $\Lambda_0=|\Lambda_0|/(1+\ii 0^+)$ (a
justification of this can be also found in the Appendix~B). Using then
$$
\frac{1}{1+u\xxsl}= \frac{1}{1- u^2 |p|^2(\Lambda_0)^{-2} } 
[1-u\psla(\Lambda_0)^{-1}]
$$
and rescaling $u|p|(\Lambda_0)^{-1}\to u(1+\ii 0^+)$ 
we see that $\cM_n$ in Eq.\  \Ref{cM} becomes\footnote{We use the different 
symbol $\tilde\cM_n$ to indicate that these numbers are obtained with a 
simplified IR regularization.} 
\eq
\label{I1}
\tilde\cM^{\vnu}_{n;\vs} = \frac{1}{8 \pi^2}
\int_{|m|}^\Lambda\dd |p| |p|^{3-n}
\cJ^{\vnu}_{n;\vs}(p) 
\eqend
where 
\eqa
\label{cJ}
\cJ^{\nu_1\ldots\nu_n}_{n;\vs} := 
\int_0^\infty \dd u u^{n-1} 
\left( \frac{1}{1- [u(1+\ii 0^+)]^2}\right)^{n+1} \nonu \times 
\left< \tras (1-u\xisl) 
\gamma_{s_1}^{\nu_1} (1-u\xisl) \cdots
\gamma_{s_n}^{\nu_n} (1-u\xisl) \right> ; 
\eqaend
we used $(2\pi)^{-4}\intML\dd^4 p \, g(p) = (8\pi^2)^{-1}
\int_{|m|}^\Lambda \dd |p| |p|^3 \left<g(|p|\xi )\right> $ with
\eq
\label{av}
\left< g( \xi ) \right> : = \, 
\frac{1}{2\pi^2} \intS \dvxi \, \delta(|\xi|-1) \, g(\xi )  
\eqend
the angular average (i.e., integration over the unit sphere in $\R^4$). 
We now see that
$\Lambda_0=|\Lambda_0|/(1+\ii 0^+)$ is needed to specify how to treat the 
singularity in the $u$-integral. These $u$-integrals are then finite 
(see Eqs.\  \Ref{cN} and \Ref{cN1} below). The result we get is 
independent of $|\Lambda_0|$, as expected. It shows explicitly 
that our expansion leads to an expansion of the action in powers of 
$\Lambda$.  We are interested in $\Lambda\to\infty$.  In 
this limit, $\tilde \cM_n\propto |\Lambda|^{n-4}$ for $n< 4$ and $\propto 
|m|^{4-n}$ for $n>4$: the former terms are divergent in the UV 
(i.e., for $\Lambda\to\infty)$, the 
latter in the IR (i.e., for $m \to 0$).  
It is precisely the `boundary case' $n=4$ which 
gives rise to the logarithmic divergence. 

This result is obtained with the simplified IR treatment is 
correct only in leading order in $\Lambda$. In Appendix~B we show 
how to do the computation without this simplification, and that  
\eq
\label{MM}
\cM_n =\tilde\cM_n +\cO\left(m^2\Lambda^{2-n} \right), \quad n>2
\eqend 
showing that the simplified IR treatment gives the correct
result for the diverging terms for all $n$ but $n=2$. For $n=2$ there
are corrections $\propto m^2\log(\Lambda/m)$ which contribute to
$\Slog$ and which we therefore have to compute exactly.

\bigskip
\noindent
{\bf 3. Computation of diverging parts of the effective action.}
We now proceed to compute the 
coefficients $\cJ_n$ Eq.\  \Ref{cJ} for those terms we are interested in, 
i.e.,  for $n=1,2,3,4$. Using Eq.\  \Ref{I1} this is
straightforward: one only needs to evaluate the integrals 
\eq
\label{cN}
\cN_{n,k}= \int_0^\infty \dd u u^{n+k-1}\left( 1-[u(1+\ii 0^+)]^2
\right)^{-n-1},  
\eqend
the angular averages $\left<\xi_{\nu_1}\cdots 
\xi_{\nu_k}\right>$, and traces of products of Dirac matrices.
The integrals in Eq.\  \Ref{cN} are (cf., e.g., Eq.\  3.251(11.) in 
\cite{Grad})
\eq
\label{cN1}
\cN_{n,k} = -(-1)^{(n-k)/2}B\left(\frac{n+k}{2} , \frac{n-k}{2}+1 \right)
\eqend
where $B(x,y)=\Gamma(x)\Gamma(y)/\Gamma(x+y)$. We only 
need
\nonueqa
\cN_{2,0} = \frac{1}{4},\quad \cN_{2,2} =  -\frac{1}{4} \\
\cN_{4,0} = \frac{1}{24},\quad \cN_{4,2} =  -\frac{1}{24}  ,\quad
\cN_{4,4} = \frac{1}{8} . 
\nonueqaend
The computation of the traces of Dirac 
matrices is simplified using the following relations
\eqa
(1-u\xisl)\gamma^{\nu} = \gamma^{\nu} (1+u\xisl) -2u\xi^{\nu},\quad 
\nu=1,2,3,4\nonu
\gamma_5 \xisl\gamma_5 =-\xisl,\quad \xisl^2 = |\xi|^2
\eqaend
which follow from Eq.\  \Ref{16b}. We also need 
\eqa
\left< 1\right> =1,\quad \left< \xi_{\mu_1}\xi_{\mu_2}\right> = 
\frac{1}{4} \eta_{\mu_1\mu_2}\qquad\qquad\qquad\nonu
\left< \xi_{\mu_1}\xi_{\mu_2} \xi_{\mu_3}\xi_{\mu_4}\right> 
=\frac{1}{24} 
\left( \eta_{\mu_1\mu_2}\eta_{\mu_3\mu_4} +
\eta_{\mu_1\mu_3}\eta_{\mu_2\mu_4} + \eta_{\mu_1\mu_4}\eta_{\mu_2\mu_3}
\right) 
\eqaend
and that the angular average for a product of an odd number of components 
$\xi_{\mu_j}$ is zero. Moreover, 
\eqa
\trac(\gamma^{\mu_1}\gamma^{\mu_2} ) 
= 4 \eta^{\mu_1\mu_2}\qquad\qquad\qquad\nonu
\trac( \gamma^{\mu_1}\gamma^{\mu_2}\gamma^{\mu_3}\gamma^{\mu_4} ) 
=4 \left( \eta^{\mu_1\mu_2}\eta^{\mu_3\mu_4} - 
\eta^{\mu_1\mu_3}\eta^{\mu_2\mu_4} + \eta^{\mu_1\mu_4}\eta^{\mu_2\mu_3}
\right) \nonu
\trac( \gamma_5
\gamma^{\mu_1}\gamma^{\mu_2}\gamma^{\mu_3}\gamma^{\mu_4} ) = 
4 \epsilon^{\mu_1 \mu_2\mu_3 \mu_4}
\eqaend
where $\epsilon^{\mu_1 \mu_2\mu_3 \mu_4}$ 
is the completely antisymmetric symbol with $\epsilon^{1234}=1$.  
Note that 
$
\trac(\gamma^s_\mu)=
\trac(\gamma^{s_1}_{\mu_1}\gamma^{s_2}_{\mu_2}\gamma^{s_3}_{\mu_3})=0  
$ always.  

It is now easy to see that 
$
\cJ^\mu_{1,s} = \cJ^{\mu_1\mu_2\mu_3}_{3;s_1s_2s_3} = 0, 
$ 
thus
\eq
S_1= S_3=0 \: . 
\eqend
The simplest non-zero terms are for $n=2$. Combining the formulas 
given above it is easy to see that 
\eqa
\cJ_{2;s_1s_2}^{\mu_1\mu_2} =  
\frac{1}{8\pi^2} A_{s_1s_2} \eta^{\mu_1\mu_2}\nonu
A_{55}=-A_{00}=1,\quad A_{50}=A_{05}=0 . 
\eqaend
Thus
\eq
\label{Stwot}
\tilde \cS_2 = \Lambda^2 \frac{1}{16 \pi^2}
\int \dvx \trac(-D^\mu D_\mu + C^\mu C_\mu ) . 
\eqend
This is a gauge invariant differential operator. When acting on $1$ 
(cf.\ Eq.\  \Ref{how}) we obtain the quadratic divergent part of the effective action 
Eq.\  \Ref{S2} which is not gauge invariant. 

As mentioned, $\tilde \cS^{(2)}$ is only the leading order contribution 
to $\cS^{(2)}$.  A more careful computation 
without the simplified IR regularization gives (see Appendix B),
\eq
\label{Stwo}
\cS_2=\tilde \cS_2 - m^2\log\left( \frac{\Lambda}{|m|} \right)
\frac{1}{8 \pi^2} \int \dvx \trac(C^\mu C_\mu ) + \ldots \eqend where
`$\ldots$' are terms which remain finite for $\Lambda\to \infty$.  We
see that the subleading term which was missed by the naive IR regularization
contributes to $\Slog$. As discussed in Section \ref{sec2}, 
this term is gauge invariant.

We now turn to the case $n=4$ which leads to the logarithmic
divergence.  All relations needed to compute the $\cJ^{\nu_1
\nu_2\nu_3 \nu_4}_{\vs}$ were listed above. The result can be written
as follows 
\eq
\label{str}
\cJ^{\nu_1 \nu_2\nu_3 \nu_4}_{\vs} = \frac{1}{3} \left(
A_{\vs}\eta^{\nu_1 \nu_2}\eta^{\nu_3 \nu_4} + B_{\vs}\eta^{\nu_1
\nu_3}\eta^{\nu_2 \nu_4} + C_{\vs}\eta^{\nu_1 \nu_4}\eta^{\nu_2 \nu_3}
+ 
D_{\vs}\epsilon^{\nu_1 \nu_2\nu_3 \nu_4}
\right) 
\eqend
where $\vs=(s_1,s_2,s_3,s_4)$. 
The numbers $A_{\vs},B_{\vs},C_{\vs},D_{\vs}$ are all given 
in Table \ref{Tab1}. (We have checked this result extensively 
using the symbolic programming language MAPLE.) 
We note that the numbers $A_{\vs},B_{\vs},C_{\vs}$ 
($D_{\vs}$)
all are real (purely imaginary) and non-zero only if an even 
(odd) number of the $s_j$ equal $5$. 

%

\begin{table*}
\caption{Parameters in Eq.\  \protect \Ref{str} where 
$\vs=(s_1,s_2,s_3,s_4)$.}
\label{Tab1}
\begin{tabular}{|c|rrrrrrrrrrrrrrrr|}
\hline
$s_1$ & $\;\;$1& $\;\;$1& $\;\;$1& $\;\;$1& $\;\;$5& $\;\;$5& $\;\;$5& $\;\;$5& 
$\;\;$1& $\;\;$1& $\;\;$1& $\;\;$1& $\;\;$5& $\;\;$5& $\;\;$5& $\;\;$5  \\
$s_2$ & 1& 1& 5& 5& 1& 1& 5& 5& 1& 1& 5& 5& 1& 1& 5& 5  \\
$s_3$ &  1& 5& 1& 5& 1& 5& 1& 5& 1& 5& 1& 5& 1& 5& 1& 5 \\
$s_4$ & 1& 5& 5& 1& 5& 1& 1& 5& 5& 1& 1& 5& 1& 5& 5& 1 \\ \hline
$A_{\vs}$ & $0$& $-2$& $-2$& $2$& $-2$& $-2$& $-2$&  $0$&  $0$&  $0$&  $0$&  $0$&  $0$&  $0$&  $0$&  $0$ \\
$B_{\vs}$ & $-2$& $2$& $2$&  $0$& $4$& $2$& $2$& $-2$&  $0$&  $0$&  $0$&  $0$&  $0$&  $0$&  $0$&  $0$  \\
$C_{\vs}$ & $2$& $-2$&  $0$& $2$& $-2$&  $0$& $-2$& $2$&  $0$&  $0$&  $0$&  $0$&  $0$&  $0$&  $0$&  $0$ \\
$D_{\vs}$ & $0$&  $0$&  $0$&  $0$&  $0$&  $0$&  $0$&  $0$& 
$-$\ii$$& $-$\ii$$& $$\ii$$& $$\ii$$& $$\ii$$& $$\ii$$& $-$\ii$$& $-$\ii$$ \\
\hline
\end{tabular}
\end{table*}

Combining these results we find
\eq
\tilde S_4 = \log\left( \frac{\Lambda}{|m|} 
\right)
\frac{1}{24\pi^2}\intS\dvx\trac[ \cP_R + \cP_{I}] ,
\eqend
where
\eqa
\cP_R=\sum_{\vs}\left( 
 A_{\vs}\eta^{\nu_1 \nu_2}\eta^{\nu_3 \nu_4} 
+ B_{\vs}\eta^{\nu_1 \nu_3}\eta^{\nu_2 \nu_4} + 
C_{\vs}\eta^{\nu_1 \nu_4}\eta^{\nu_2 \nu_3} 
\right)
D^{s_1}_{\nu_1}D^{s_2}_{\nu_2}D^{s_3}_{\nu_3}D^{s_4}_{\nu_4}
\nonu
\cP_{I}=\sum_{\vs} D_{\vs} \epsilon^{\nu_1 \nu_2\nu_3 \nu_4}
D^{s_1}_{\nu_1}D^{s_2}_{\nu_2}D^{s_3}_{\nu_3}D^{s_4}_{\nu_4} 
\qquad\qquad\qquad\qquad
\eqaend
with the coefficients given in Table \ref{Tab1}. 
$\cP_R$ is a sum of 19 non-zero terms. We now claim that 
it is possible to write$\cP_R=\cP_{R,1}+\cP_{R,2}$ where 
\eqa
\label{FF}
\cP_{R,1} = -[D^\mu , D^\nu][D_\mu ,D_\nu] - [C^\mu , C^\nu][C_\mu 
,C_\nu] + [D^\mu , D^\nu][C_\mu ,C_\nu] \nonu  + [C^\mu , C^\nu][D_\mu ,D_\nu] 
+ 2 [D^\mu , C^\nu][D_\mu ,C_\nu] + 2 [D^\mu , C^\nu][C_\mu ,D_\nu] 
\eqaend
and 
\eq
\cP_{R,2} = \ii[ D^\mu,J_\mu] + \left[[D^\mu , D^\nu],[C_\mu ,C_\nu]\right] -
2[C^\mu,[D_\mu , D_\nu]C^\nu] 
\eqend
with $J_\mu$ given in Eq.\  \Ref{J}. 
Similarly, 
\eqa
\label{cpi}
\cP_{I} = \frac{\ii }{2}\epsilon^{\nu_1 \nu_2\nu_3 \nu_4}
\left[ [D_{\nu_1},D_{\nu_2}]+[C_{\nu_1},C_{\nu_2}],[D_{\nu_3},C_{\nu_4}]  
\right] . 
\eqaend
(The proof of Eqs.\  \Ref{FF}--\Ref{cpi} are straightforward calculation which we skip.)

We see that, $\cP_{R,1}$ equals $\half
(\cF^+_{\mu\nu}(\cF^+)^{\mu\nu}+\cF^-_{\mu\nu}(\cF^-)^{\mu\nu}) $ with
$\cF^\pm_{\mu\nu}$ defined in Eq.\ \Ref{cFpm}.  The remaining terms
are linear combinations of commutators!  Using the cyclicity of the
matrix trace we thus obtain \eq \trac \cP_{R,2} = \partial^\mu \trac
J_\mu, \quad \trac \cP_{I} = 0 .  \eqend This implies Eqs.\
\Ref{main}--\Ref{J} and completes our computation. \QED

\bigskip
\noindent {\em Remark:} Note that $\cP_R$ and $\cP_I$ are not
differential operators but polynomials (i.e., there are no terms
$(\cdots)D_\mu$).  This implies that both these terms are gauge
covariant which, as we believe, is remarkable.  

\section{Conclusions}
The regularization which we used was simple but not manifestly gauge
invariant. For the result computed in this paper the latter property
is irrelevant: since the logarithmic divergence is regularization
dependent one can compute it using any regularization. However, we
believe that our method is useful even for computing the finite part
of the effective action, i.e., $S^{(0)}(A)$ in Eq.\ \Ref{S0}. We
stress again that the simplified IR regularization used in the main text
is not appropriate in this computation but the formulas given in
Appendix B should be used. We conjecture that $S^{(0)}(A)$ computed in
this way is gauge invariant.

As mentioned in the Remark at the end of Section \ref{sec3}, we
defined a renormalized trace $\TR_{|\oDn|}$ using the free Dirac
operator $\oDn$.  More general we could use the Dirac operator $\oDB$
with some fixed non-trivial Yang-Mills field $B$. In particular, we
expect that the standard $\zeta$-function regularization of the
logarithm of the determinant of $\oDA$ should be identical with
$$
\TR_{|\oDA|}  \log \left(\frac{\oDA+\ii m
}{\Lambda_0}\right)
$$
were the regularization function is $f(t)=\exp(-t^2)$.  The latter
definition has the advantage that it is manifestly gauge invariant,
but it seems less easy to use for explicit computations as ours.  It
is natural to expect that the difference between the latter definition
and $S^{(0)}(A)$ in Eq.\ \Ref{S0} is also proportional to 
$\Slog(A)$.

Effective action computations are used in many applications of quantum
field theory. We believe that the methods which we presented
should be useful in other such contexts as well.

\bigskip

\noindent{\bf Acknowledgment:} I would like to thank A.\ Laptev, J.\
Mickelsson, S.\ Paycha, F.\ Scheck and K.\ Wojciechowski for their
interest and helpful discussions and S.\ Paycha for comments on the
manuscript.  This work was supported by the Swedish Natural Science
Research Council (NFR).

\section*{Appendix A: More on regularized traces} 
\app In this Appendix we outline elementary proofs of some facts about
regularized traces stated in the main text.

\bigskip
\noindent {\bf The logarithmic divergence.} 
We compute the regularized trace in Eq.\ \Ref{TrLam1} for
an operator $a$ with a symbol allowing for an asymptotic expansion
as in Eq.\ \Ref{as0}. It is easy to see that 
the contribution of $\sigma_{k}[a](p,x)$ to $\TraL(a)$ is 
\nonueqa \int_0^\infty d|p| |p|^{k+3} f(\pL) \intS\dvxi \, 
\delta(|\xi|-1) \intS\dvx\, \tra{\sigma_k[a](\xi ,x)}\nonueqaend
where we used the homogeneity of $\sigma_k[a]$.  
Changing variables, $|p|\to u=|p|/L$, and comparing with Eq.\
\Ref{TrReg} we see that for all $k\geq -3$, \eq c_{k+4}(a) =
N_{k} \intS\dvxi\, \delta(|\xi|-1) \, \tra{\sigma_{k}[a](\xi ,x)} \eqend with
$N_{k}=\int_0^\infty \dd u u^{k+3} f(u) $ constants depending on $f$.
For $k=-4$ the computation above does not make sense (the constant
$N_{-4}$ diverges), but we can compute $c_{\rm 
log}(a)$ as follows.  We first subtract from the symbol of $a$ the 
diverging part which we already accounted for and define, 
\eq
\sigma^\perp_{-3}[a](p,x) \ := \ \sigma[a](p,x) - \sum_{j= 0 
}^{K+3} \sigma_{K-j}[a] (p,x) = \sigma_{-4}[a] (p,x) + \cO(|p|^{-5}) 
\: . 
\eqend
Eq.\ \Ref{TrReg} then suggests that $$c_{\rm log}(a) = 
\lim_{\Lambda\to\infty} \frac{1}{\log(\Lambda)} \intS\dvp f(\pL) 
\intS\dvx\tra\, \sigma[a]^\perp_{-3}(p,x) \: .  $$ 
Computing this using L'Hospital's rule we obtain 
\nonueqa c_{\rm 
log}(a) = \lim_{\Lambda\to\infty}\Lambda \intS\dvp 
f'(\pL)(-\mbox{$\frac{|p|}{\Lambda^2}$}) 
\intS\dvx\tra\, \sigma[a]^\perp_{-3}(p,x) \\
= \lim_{\Lambda\to\infty} \left( \intS\dvp (-f'(\pL) \pL) 
\intS\dvx\tra\, \sigma[a]_{-4}(p,x) +\cO(\Lambda^{-1}) \right) 
\: .  \nonueqaend 
Changing variables etc.\ as above and using $ \int_0^\infty du (-f'(u) )  = 
f(0)=1$ (independent of $f$!)  we obtain 
\eq 
\label{Wres1} 
c_{\rm log}(a) = 
\intS\dvxi\, \delta(|\xi|-1) \intS\dvx\tra{\sigma_{-4} (a)(\xi ,x)} . 
\eqend 
Recalling Eq.\ \Ref{Wres_def} we obtain Eq.\ \Ref{aa}. \QED

\bigskip
\noindent {\bf Renormalized traces.} It is obvious that changing the
regularization functions $f(t)\to \tilde f(t)=f(t/s)$ for some fixed
$s>0$, amounts to changing $\Lambda\to s \Lambda$, and thus changes
$c^{(0)}\to c^{(0)} + \log(s)\, c_{\rm log}$. Thus \Ref{TrReg} is
obvious for this special case.  For more general changes $f(t)\to
\tilde f(t)$ of the regularization function, Eq.\ \Ref{TrReg} can
be shown using
$$
\intS\dvp f(\pL) \intS\dvx\tra\, \sigma[a]^\perp_{-3}(p,x) =
c_{\log}(a) \log\left(\frac{\Lambda}{|m|}\right) + c^{(0)}(a) + {\cal
O}(\Lambda^{-1}) 
$$
which follows from our discussion above. 

\section*{Appendix B: Computation details}
\app In this Appendix we present some details concerning our
computations discussed in the main text. In particular, we give
explicit formulas for the Dirac matrices, and we also show show how to
compute the structure constants $\cM_n$ in Eq.\ \Ref{cM} exactly,
i.e., without the simplified IR regularization. We also prove the
Lemma in Section~4.1 and Eq.\ \Ref{MM}, and we give some details about
the computation yielding Eq.\ \Ref{Stwo}.

\subsection*{B.1. Dirac matrices}
A convenient representation for the Dirac matrices is as follows,
\eqa
\gamma^j=\left( \bma{ll} \noll&\sigma_j\\ \sigma_j&\noll\ema\right), 
\quad j=1,2,3, \quad  
\gamma^4=\left( \bma{rr} \noll&\ii \ett \\ -\ii\ett &\noll\ema\right), 
\quad 
\gamma_5=\left( \bma{rr} \ett& \noll \\ \noll  & -\ett \ema\right)\;  
\eqaend
where $\ett$ and $\noll$ are the $2\times 2$ unit- and zero matrices
and
$$
\sigma_1=\left( \bma{rr}0&1\\ 1&0 \ema\right)\; , 
\quad 
\sigma_2=\left( \bma{rr}0&-\ii\\ \ii &0 \ema\right)\; , 
\quad 
\sigma_3=\left( \bma{cc}1&0\\ 0&-1 \ema\right)\; , 
$$
the Pauli sigma matrices as usual.

\subsection*{B.2. Details about the gradient expansion}
We start by rewriting the $\cM_n$ in a convenient form.  We define 
\eq
P_{\eps}:=\frac{1}{2}\left( 1 +\eps \frac{\psla}{|p|} \right) ,\quad
\eps=\pm 
\eqend
 which are orthogonal projections, $P_\eps P_{-\eps}=0$ and
$P_\eps^2=P_\eps$, satisfying $P_+ +P_-=1$. We then can write
$$
(1-u\xxsl)^{-1}=\sum_{\eps=\pm} P_\eps\frac{1}{1+
u\frac{\eps|p|+\ii m}{\Lambda_0} }
$$
which we insert $n+1$ times in Eq.\  \Ref{cM},
\nonueqa
\cM^{\nu_1\ldots\nu_n}_{n;s_1\ldots s_n} =\, (\Lambda_0)^{-n}
\intL\dvp \int_0^\infty \dd u u^{n-1}  \nonu \times
\sum_{\eps_1,\ldots,\eps_{n+1}=\pm } \left( \prod_{j=1}^{n+1} 
\frac{1}{1+u\frac{\eps_j |p|+\ii m}{\Lambda_0} }\right) 
\tras \left( P_{\eps_1} 
\gamma_{s_1}^{\nu_1} P_{\eps_2} \cdots
\gamma_{s_n}^{\nu_n}  P_{\eps_{n+1} } \right) .
\nonueqaend
We thus obtain 
\eq
\label{A2}
\cM^{\nu_1\ldots\nu_n}_{n;s_1\ldots s_n} = \sum_{\eps_1,\ldots,\eps_{n+1}=\pm }
\cI_{n;\eps_1,\ldots,\eps_{n+1}}\tras \left< P_{\eps_1} 
\gamma_{s_1}^{\nu_1} P_{\eps_2} \cdots
\gamma_{s_n}^{\nu_n}  P_{\eps_{n+1} } \right> 
\eqend
with 
\eq
\label{A3}
\cI_{n;\eps_1,\ldots,\eps_{n+1}}=\cI_{n;k} \;,\quad  \mbox{ $k$ such that } 
\sum_{j=1}^{n+1} \eps_j = n+1-2k 
\eqend
and
$$
\cI_{n;k} =(\Lambda_0)^{-n}\frac{1}{8\pi^2}\int_0^\Lambda\dd|p| |p|^3
\int_0^\infty \dd u u^{n-1} \left(\frac{1}{1+u\frac{|p|+\ii
m}{\Lambda_0} }\right)^{n+1-k}\left(\frac{1}{1+u\frac{-|p|+\ii
m}{\Lambda_0} }\right)^k \: .
$$
Rescaling $u \Lambda /\Lambda_0\to u$ and introducing 
$\xi=|p|/\Lambda$ yields 
\eq
\label{cI}
\cI_{n;k} =(\Lambda)^{4-n}\frac{1}{8\pi^2}\int_0^1\dd \xi \xi^3 
\int_0^\infty \dd u u^{n-1}
\left(\frac{1}{1+u[\xi+\ii \frac{m}{\Lambda}] }\right)^{n+1-k}
\left(\frac{1}{1+u[-\xi +\ii \frac{m}{\Lambda}] }\right)^k \: . 
\eqend
\bigskip

\noindent{\bf Proof of the Lemma in Section~4.1.}  We note that
$$\tras \left< P_{\eps_1} \gamma_{s_1}^{\nu_1} P_{\eps_2} \cdots
\gamma_{s_n}^{\nu_n} P_{\eps_{n+1} } \right> =\,
:T^{\vnu}_{\veps,\vs}$$ is invariant under $\eps_j\to -\eps_j$ (since
the latter transformation amounts to the variable change $\xi\to -\xi$
in the integral Eq.\ \Ref{av} defining the angular average). Moreover,
the cyclicity of trace and $\gamma_5^2=1$ implies that
$T^{\vnu}_{\veps,\vs}$ does not change if we replace all $P_{\eps_j}$
and $\gamma_{s_j}^{\nu_j}$ by $\gamma_5 P_{\eps_j}\gamma_5$ and
$\gamma_5\gamma_{s_j}^{\nu_j}\gamma_5$, respectively.  Using $\gamma_5
P_\eps \gamma_5=P_{-\eps}$ and $\gamma_5 \gamma_s^\nu \gamma_5 = -
\gamma_s^\nu$ we obtain $ T^{\vnu}_{\veps,\vs} = (-1)^n
T^{\vnu}_{-\veps,\vs}, $ and using
$T^{\vnu}_{-\veps,\vs}=T^{\vnu}_{\veps,\vs}$ this proves that
$T^{\vnu}_{\veps,\vs}$ --- and thus $\cM_n$ in Eq.\ \Ref{cM} --- is
non-zero only for even $n$.

From Eq.\ \Ref{A3} it is obvious that $\eps_j\to -\eps_j$ corresponds
to to $k\to n+1-k$, and thus
$T^{\vnu}_{\veps,\vs}=T^{\vnu}_{-\veps,\vs}$ implies that we can
replace $\cI_{n;k}$ by $[\cI_{n;k}+\cI_{n;n+1-k}]/2$ in Eq.\
\Ref{A2}. We can write the latter as a sum of the terms which are even
and odd under the change the sign of the mass $m\to -m$. A simple
change of variables shows that $u$-integrals in the odd term
$$
\frac{1}{4}\left[ \cI_{n;k}(m) +\cI_{n;n+1-k}(m) -
\cI_{n;k}(-m)-\cI_{n;n+1-k}(-m) \right]
$$ 
can be written as follows ($n$ even), 
$$
\frac{1}{4} \int_{-\infty}^\infty \dd u u^{n-1} 
\left(\frac{1}{1+u[\xi+\ii \frac{m}{\Lambda}] }\right)^{n+1-k}
\left(\frac{1}{1+u[-\xi +\ii \frac{m}{\Lambda}] }\right)^k  
$$
plus the same integral but with $k$ and $n+1-k$ interchanged.  The
latter integrals can be computed using Cauchy's theorem: the poles of
the integrand are in $u=-1/(\xi +\ii m/\Lambda)$ and $u=1/(\xi -\ii
m/\Lambda)$ and thus both always in the same half of the complex $u$-plane
(upper or lower, depending on the sign of $m$). Computing the
integral by closing the integration path in the half plane where the
integrand is analytic (which is possible since the integrand vanishes
like $\cO(|u|^{-2})$ for $|u|\to \infty$) one sees that the integral
is zero. This implies $\cM_n(-m)=\cM_n(m)$. \QED

\bigskip

\noindent{\bf Proof of Eq.\ \Ref{MM}:} Our discussion above implies
that we can replace $\cI_{n;k}$ in Eq.\ \Ref{A2} by $\Re \cI_{n;k} = [
\cI_{n;k}(m) + \cI_{n;k}(-m) ]/2$.

We are interested in the terms which diverge for $\Lambda\to\infty$.
To isolate them it is convenient to determine 
$\partial\cM_n/\partial\Lambda$. We thus compute 
\eq
\label{dcI}
\frac{\partial}{\partial\Lambda }\Re \cI_{n;k}
= \Lambda^{3-n} \frac{1}{8\pi^2}I_{n,k}(\mL)
\eqend
where we introduced the functions
\eq 
\label{Ink}
I_{n,k}(\eta) = \Re \int_0^\infty \dd u u^{n-1}
\left(\frac{1}{1+u[1 +\ii\eta ]}\right)^{n+1-k}
\left(\frac{1}{1+u[-1+\ii \eta ] }\right)^k .  
\eqend 
Note that the functions $I_{n,k}(\eta)$ are well-defined for all real
$\eta\neq 0$, have a finite limit $I_{n,k}(0^+)$ as $\eta\to
0$, and they have series expansions in $\eta^2$.\footnote{To see
this note that $I_{n,k}(\eta) = \Re \int_0^\infty \dd s \left(\frac{1}{s+1
+\ii\eta }\right)^{n+1-k} \left(\frac{1}{s -1+\ii \eta }\right)^k $.}

It is easy to see that with the simplified regularization used in the
main text we can obtain a formula for
$\tilde\cM^{\nu_1\ldots\nu_n}_{n;s_1\ldots s_n}$ as in Eqs.\
\Ref{A2}--\Ref{A3} but with $\cI_{n;k}$ replaced by 
\eq
\tilde\cI_{n;k} = 
\frac{1}{8\pi^2}\int_0^\Lambda\dd|p| |p|^{3-n}
I_{n;k}\left( 0^+ \right) .  
\eqend 
We thus get 
\eq
\label{A4}
\frac{\partial}{\partial\Lambda}\left(\cI_{n;k} - \tilde \cI_{n;k}
\right) = \frac{1}{8\pi^2}\Lambda^{3-n} \left( 
I_{n;k} (\mL) -I_{n;k}(0^+) \right) 
= \cO(m^2 \Lambda^{1-n}) , 
\eqend
which proves Eq.\ \Ref{MM}. \QED

\bigskip
\noindent {\em Remark:} We now can explain the reason for our choice
$\Lambda_0=|\Lambda_0|/(1+\ii 0^+)$ in the main text: this yields a
regularization specifying the otherwise undefined integrals
$I_{n;k}(0)$, and from Eq.\ \Ref{A4} it is clear that this is the
regularization yielding a result identical with the one obtained with
the proper regularization, up to lower order terms.
\bigskip

\noindent {\bf Computation of $\cS_2$.} 
For $n=2$ we need compute $\cM_2$ in
Eq.\ \Ref{cM} exactly, using the formulas given above.
 
Similarly as explained in the main text we compute (cf.\ Eq.\ 
\Ref{A2})
$$
\tras \left< P_{\eps_1} 
\gamma_{s_1}^{\nu_1} P_{\eps_2} \cdots
\gamma_{s_2}^{\nu_2} P_{\eps_3}  \right> = 
\delta_{\eps_1,\eps_3}
\eta^{\nu_1\nu_2}\left( 1-\eps_1\eps_2(-1)^{s_1}  \right) .
$$
Moreover, the integrals defined in
Eq.\ \Ref{Ink} for $n=2$ and $k=0,1$ are, \nonueqa I_{2,0}(\eta) &=& \Re
\frac{1}{2(1+\ii\eta)^2} =\frac{1}{2}- \frac{3}{2}\eta^2+\cO(\eta^4)\\
I_{2,1}(\eta) &=& \Re \frac{1}{4(1+\ii\eta)}\left( (1+\ii\eta)
\log\left(\frac{1+\ii\eta}{1-\ii\eta}\right) - 2 \right) = \frac{1}{2}
+ \frac{1}{2}\eta^2+\cO(\eta^4), \nonueqaend and with Eqs.\ \Ref{dcI},
\Ref{A2} and \Ref{A3} we can compute
$\partial\cM_2/\partial\Lambda$. Straightforward computations yield
\eqa 
\label{cMtwo}
\cM^{\nu_1\nu_2}_{2;s_1s_2} = \delta_{s_1s_2} \eta^{\nu_1\nu_2}
\frac{1}{16\pi^2} \left(\Lambda^2 A_{s_1s_2} + m^2\log\left(
\frac{\Lambda}{|m|}\right) A^{(0)}_{s_1s_2} +\cO(\Lambda^{0}) \right)
\nonu A_{55}=-A_{00}=1, \quad A^{(0)}_{55}=-2,\quad A^{(0)}_{00}=0 ,
\eqaend
and with Eq.\ \Ref{ex} we obtain Eqs.\ \Ref{Stwot}--\Ref{Stwo}.

%

\end{document}